\documentclass[12pt,fleqn]{article}
\usepackage{amsmath,amssymb}
\usepackage{bm}
\usepackage[dvips]{graphicx}
\allowdisplaybreaks[3]

\newcommand{\be}{\begin{equation}}
\newcommand{\ee}{\end{equation}}
\newcommand{\ba}{\begin{eqnarray}}
\newcommand{\ea}{\end{eqnarray}}

\newcommand{\rw}{\rightarrow}
\newcommand{\De}{\Delta}

\newcommand{\Om}{\Omega}

\begin{document}

\title{Baryonic Resonances from the Interactions of the Baryon Decuplet and Meson Octet }

\author{E. Oset, S. Sarkar and M. J. Vicente Vacas}

\maketitle 

\begin{center}
\textit{Departmento de F\'isica Te\'orica and IFIC,
Centro Mixto Universidad de Valencia-CSIC,
Institutos de Investigaci\'on de Paterna, Aptd. 22085, 46071
Valencia, Spain}\\
\end{center}

\vspace{0.4cm}

\abstract{
 We study $S$-wave interactions of the baryon decuplet with the octet of
pseudoscalar mesons using the lowest order chiral Lagrangian.
We find two bound states in the $SU(3)$
limit corresponding to the octet and decuplet representations. These are found
to split into eight different trajectories in the complex plane when the $SU(3)$
symmetry is broken gradually. Finally,
we are able to provide a reasonable description for a good number of 4-star
${\frac{3}{2}}^-$ resonances listed by the Particle Data Group. In particular,
the $\Xi(1820)$, the $\Lambda(1520)$ and the $\Sigma(1670)$ states are 
well reproduced. We predict a few other resonances and also evaluate the couplings of the observed resonances to
the various channels from the residues at the poles of the scattering matrix from
where partial decay widths into different channels can be evaluated.}

%%%%%%%%%%%%%%%%%%%%%%%%%%%%%%%%%%%%%%%%%%%%
%% MAINMATTER
%%%%%%%%%%%%%%%%%%%%%%%%%%%%%%%%%%%%%%%%%%%%
\section{Introduction}

The introduction of unitary techniques in a chiral dynamical treatment of the
meson baryon interaction has been very successful. It has lead to good
reproduction of meson baryon data with a minimum amount of free parameters, and
has led to the dynamical generation of many low lying resonances which qualify
as quasibound meson baryon states~\cite{kaiser,ramos,oller}. In particular,
the application of these techniques to the $s$-wave 
scattering of the baryon octet and the pseudoscalar meson octet have led to the
successful description of many $J^P=\frac{1}{2}^-$ resonances like the
$N^*(1535)$, the $\Lambda(1405)$, the $\Lambda(1670)$, the $\Sigma(1620)$ and
the $\Xi(1620)$~\cite{inoue,jido,manolo,carmen,bennhold}. Naively one may expect that this scheme is not suitable for
studying $d$-wave resonances due to a large number of unknown parameters in
the corresponding chiral Lagrangian. However $d$-wave resonances could be 
studied in $s$ wave interactions of the meson octet with the baryon
decuplet~\cite{lutz}, in
which case chiral dynamics is quite predictive. The fact that some well known
$d$ wave resonances like the $N^*(1520),\,N^*(1700),\,\Delta(1700)$ have very
large branching ratios to the $N\pi\pi$ channel though the $N\pi$ channel is
favoured by phase space, lends support to this scheme.    

\section{Baryon Decuplet-Meson Octet Interaction: Results and Discussion}

  We have performed a systematic study~\cite{sarkar1} of the $s$-wave interaction of the baryon decuplet
with the meson octet taking the dominant lowest order chiral Lagrangian, which
accounts for the Weinberg Tomozawa term. 
In this talk we will present the main results of this work. 

\begin{figure}
\includegraphics[width=0.43\textwidth]{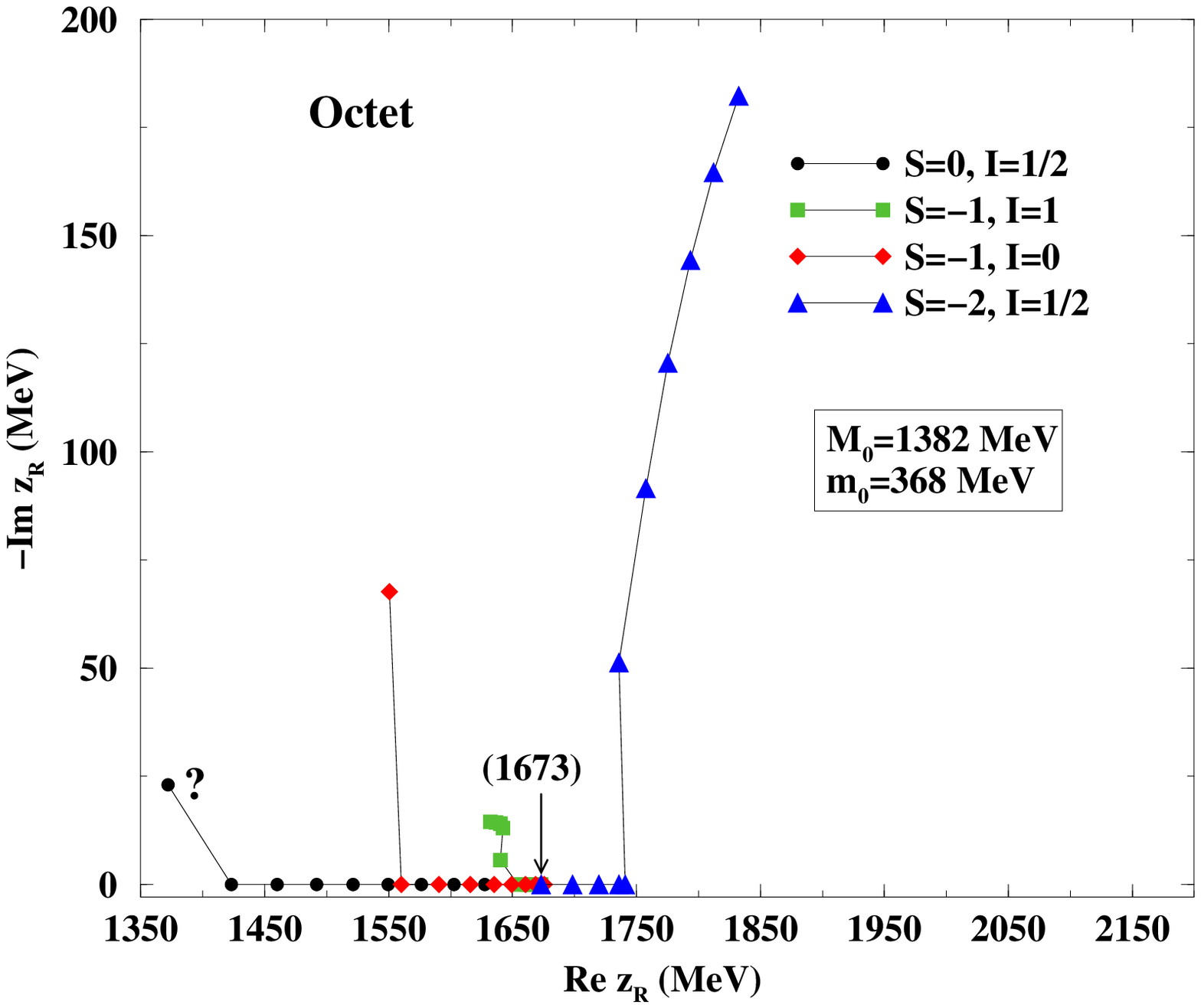}\hspace*{1cm}
\includegraphics[width=0.43\textwidth]{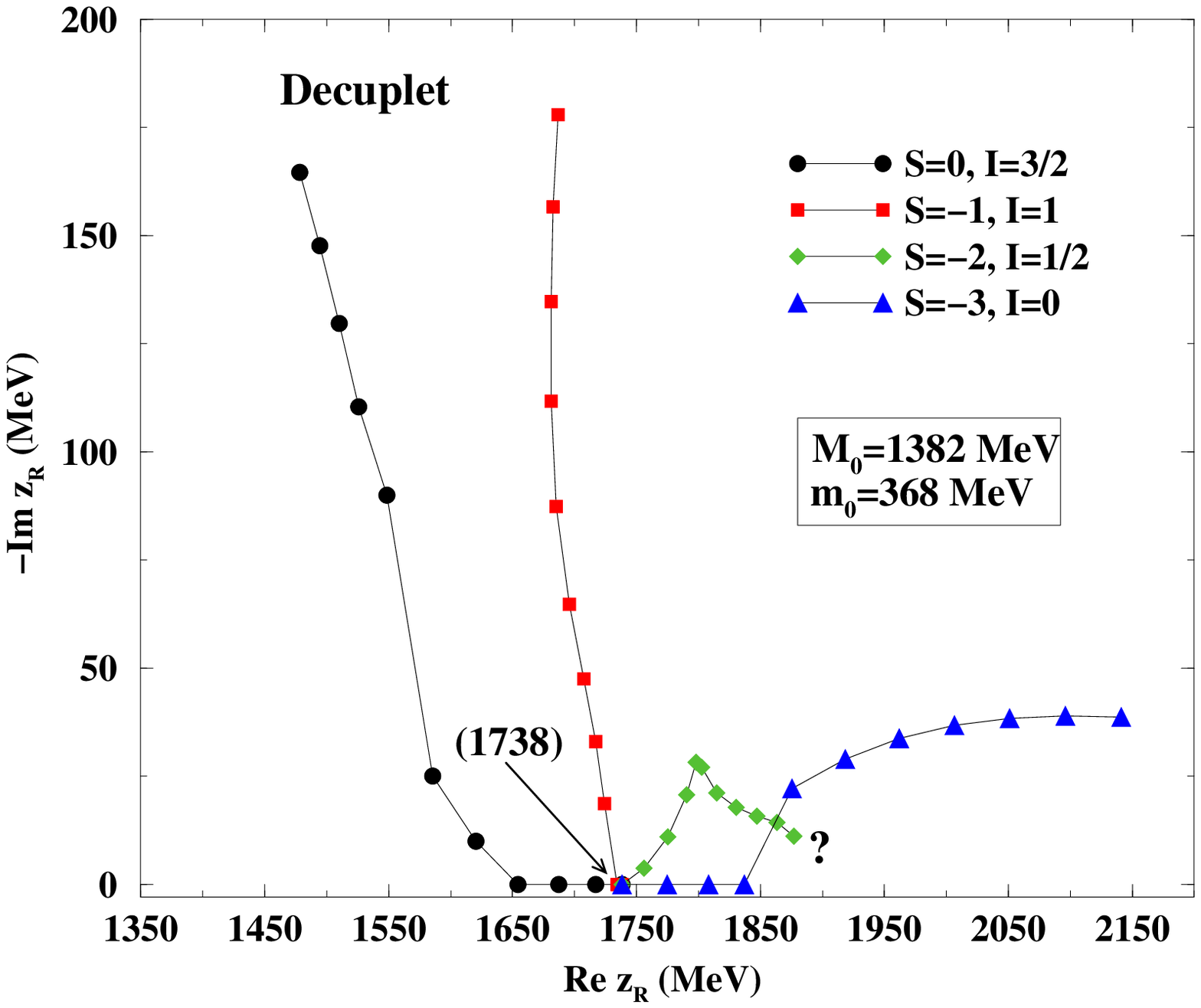}
\caption{Trajectories of the poles in the scattering amplitudes for different
values of the $SU(3)$ breaking parameter.}
\label{trajfig}
\end{figure}

The tree-level scattering amplitude involving the baryon decuplet and the
pseudoscalar octet  
%\be
%V_{ij}=-\frac{1}{4f^2}C_{ij}(k^0+k^{\prime 0}).
%\label{poten}
%\ee
is used as the kernel of the Bethe Salpeter equation  to
obtain the transition matrix fulfilling exact unitarity 
in coupled channels. In this approach, the only free parameter is the
subtraction constant in the dimensionally regularized meson baryon loop function
for which we took a natural size value. 
We have looked in detail at the $\frac{3}{2}^-$ 
resonances which are generated dynamically by this interaction, by searching for
poles of the transition matrix in the complex plane in different Riemann sheets.
 The search was done
systematically, starting from an $SU(3)$ symmetric situation where the masses of
the baryons are made equal and the same is done with the masses of the mesons. In this
case we found attraction in an octet, a decuplet and the 27 representation,
while the interaction was repulsive in the 35 representation.  In the $SU(3)$
symmetric case all states of the $SU(3)$ multiplet are degenerate and the
resonances appear as bound states with no width. As we gradually break $SU(3)$
symmetry by changing the masses, the degeneracy is broken and the states with
different strangeness and isospin split apart generating pole trajectories in the
complex plane which
lead us to the  physical situation in the final point, as seen in
fig.~\ref{trajfig}.  This systematic search
allows us to trace the poles to their $SU(3)$ symmetric origin, although there is
 a mixing of representations when the symmetry is broken. In addition we also
 find poles which only appear for a certain amount of symmetry breaking and thus
have no analog in the symmetric case. 

\begin{figure}
\includegraphics[width=1.0\textwidth]{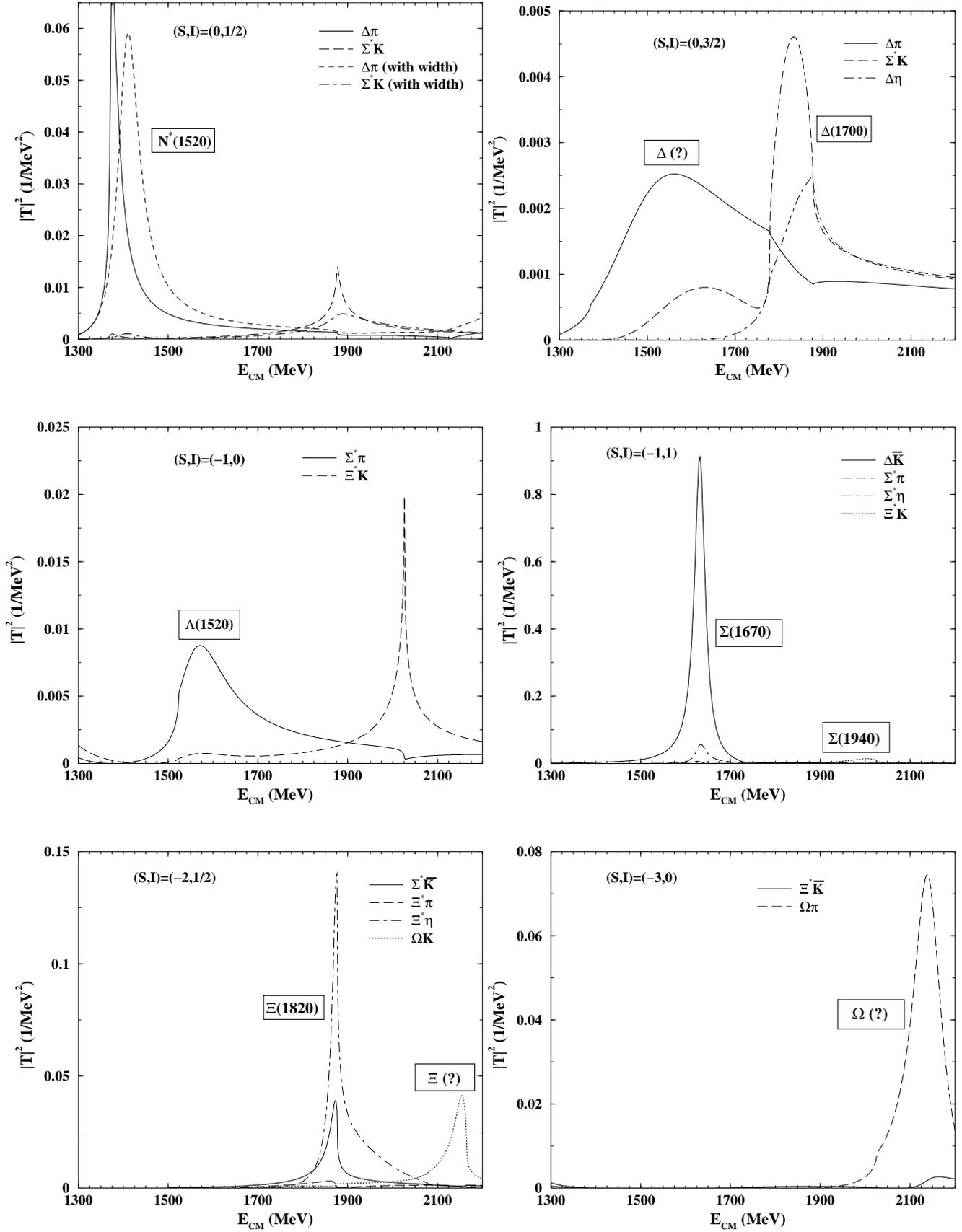}
\caption{Amplitudes for $3/2^-$ resonances generated dynamically in decuplet
baryon - pseudoscalar octet interactions.}
\label{allfigs}
\end{figure}

 The search for poles is an ideal technique to identify resonances in addition
 to visual inspection of the amplitudes. On the other hand, in ref. \cite{lutz} the
 speed plot technique is used to identify resonances. However, the speed plot,
 involving derivatives of the amplitudes,
 produces artificial peaks at the thresholds  of new channels, because the cusp 
 of the amplitudes,  and should not be used there.  Neglecting this fact, some
 peaks which are actually reflections of cusps in the speed plot are
 misidentified as resonances in \cite{lutz} while we do not find poles, not even
  making reasonable changes in the input data.
 
  We have also evaluated the residues of the poles from where the couplings of the
resonances to the different coupled channels were found and this allowed us to
make predictions for partial decay widths into a decuplet baryon  and a meson.
There is very limited experimental information on these decay channels but,
even then, it represents an extra check of consistency of the results  which
allowed us to more easily identify the resonances found with some resonances
known, or state that the resonance should correspond to a new resonance not yet
reported in the Particle Data Book (PDB). In particular, in view of the information of
the pole positions and couplings to channels we could associate some of the
resonances found to the $N^*(1520)$, $\De(1700)$, $\Lambda(1520)$, $\Sigma(1670)$,
$\Sigma(1940)$, $\Xi(1820)$ resonances tabulated in the PDB. 
We could also favour the
correspondence of a resonance found in $S=-3,\ I=0$ to the $\Om(2250)$ on the basis
of the quantum numbers, position and compatibility of the partial decay width found
with the total width of the $\Om(2250)$. The scattering amplitudes for different
values of strangeness and isospin as a function of the C.M. energy are shown in
fig.~\ref{allfigs}. 

We also found several extra resonances, well identified by poles in the complex
plane which do not have a correspondent one in the PDB. Some of them are too 
broad,
which could justify the difficulty in their observation, but two other resonances,
the $\De(1500)$ with a width around 300 MeV and the $\Xi(2160)$ with a width of
about 40 MeV stand much better chances of observation. The first one because of its
large strength in the $\De\pi$ channel and the second one because of its
narrowness. 

In addition, our study produces couplings of the resonances to baryon-meson
channels which could facilitate the identification when further information on
these branching ratios is available.
We have found that in some cases the dynamically generated resonances couple
very strongly to some channels with the threshold above the resonance energy,
which makes them qualify as approximately single channel quasibound meson-baryon
states.

%The experimental implications of this work are important since clear predictions
%on the partial decay widths into a baryon of the $\De$ decuplet and a meson of
%the pion octet are made, which are amenable to experimental observation.

Lastly, we mention the very interesting case concerning the $\De K$
resonance~\cite{sarkar2} 
found in $S=1,\ I=1$ which
shows up as a pole in the unphysical Riemann sheet and leads to $\De K$ cross sections much larger than
the corresponding $\De K$ cross sections in $I=2$. We suggest that
these two cross sections
could be seen in $\De K$ production in $pp\rw\Lambda\De K$ and $pp\rw\Sigma\De
K$ reactions and could provide evidence for this new 'exotic pentaquark'
state, although its structure is more efficiently taken into account in the
meson baryon picture where it has been generated. This state, however, appears
as a broad bump above the $K \Delta$ threshold and its strength is also tied to
details of the interaction. Hence, the ultimate fate of this resonance is
difficult to predict at the present stage.\\

{\bf Acknowledgments:}\ \ 
This work is partly supported by DGICYT contracts BFM2002-01868, 
and the E.U. EURIDICE network contract HPRN-CT-2002-00311.
This research is part of the EU Integrated Infrastructure Initiative
Hadron Physics Project under contract number RII3-CT-2004-506078.

\vfill
\end{document}